\documentclass[twocolumn,aps,pre]{revtex4}
\usepackage{makeidx}
\usepackage{amsmath}
\usepackage{amssymb}
\usepackage{graphicx}
\usepackage{times}

\begin{document}

\title{\sffamily\bfseries\large Universal distribution of threshold forces
          at the depinning transition}
\author{\sffamily\bfseries\normalsize Andrei A. Fedorenko, Pierre Le
Doussal and Kay J\"org Wiese}

\affiliation{CNRS-Laboratoire de Physique Th{\'e}orique de l'Ecole
Normale Sup{\'e}rieure, 24 rue Lhomond, 75231 Paris Cedex, France}
\date{July 10, 2006 }

\begin{abstract}
We study the distribution of threshold forces at the depinning
transition for an elastic system of finite size, driven by an external
force in a disordered medium at zero temperature. Using the functional
renormalization group (FRG) technique, we compute the distribution of
pinning forces in the quasi-static limit. This distribution is
universal up to two parameters, the average critical force, and its
width. We discuss possible definitions for threshold forces in
finite-size samples. We show how our results compare to the
distribution of the latter computed recently within a numerical
simulation of the so-called critical configuration.
\end{abstract}

\maketitle

\section{Introduction} The dynamics of elastic objects driven by an
external force in disordered media has attracted considerable
theoretical and experimental interest during the last years
\cite{fisher-phys-rep98,kardar-phys-rep98,brazovskii03}.  The reason
is twofold. On one hand, elastic objects in disordered media exhibit
the rich behavior of glassy systems and thus their study can help us
to understand the physics of more complex systems such as spin
glasses~\cite{binder-86} or random field systems \cite{random-fields}.
On the other hand, the motion of elastic objects in disordered media
is an adequate description of many experimental systems. One can
divide these systems into two classes.  The first class comprises
periodic systems, the most prominent examples being charge density
waves (CDW) in solids.  These start sliding when the applied electric
field becomes large enough \cite{cdw}. Vortex lines in disordered
superconductors form a quasi-ordered periodic Brag glass phase
\cite{vortex,bragg}.  The second class includes propagating interfaces
such as domain walls in magnetically or structurally ordered systems
\cite{domain-walls-exp}, interfaces between inmiscible fluids in
porous media \cite{fluids-invasion} or dislocation lines in metals
\cite{dislocation}. To unify the mathematical description of these
different systems one uses the notion of ``manifolds''.  In all these
systems the interplay between quenched disorder and elasticity leads
to a complicated response of the system to an applied external force.
At zero temperature, a driving force $f$ exceeding a certain threshold
value $f_c$ is required to set the elastic manifold into motion. This
depinning transition shares many features with critical phenomena
\cite{fisher85}: characteristic lengths diverge close to the
transition as $\xi \sim (f-f_c)^{-\nu}$ and the system becomes
extremely sensitive to small perturbations.  Following the description
of standard critical phenomena, one can identify the ordered phase
with the moving phase with force $f>f_c$, and the order parameter with
the velocity $v$ which vanishes as $v\sim (f-f_c)^{\beta}$ at the
transition. One also introduces the dynamic exponent $z$ which relates
time and space by $t\sim x^z$. The critical force $f_c$, which must be
tuned to reach the scale-invariant regime, plays a role similar to the
critical temperature in thermal phase transitions. There are many
subtleties however within this analogy, since depinning is a
non-equilibrium transition at zero temperature, where quenched
disorder dominates.  As the corresponding static problem of elastic
manifolds in disordered media \cite{fisher86}, the depinning problem
suffers from two peculiarities, when compared to standard critical
phenomena: First an infinite set of operators becomes relevant
simultaneously below the internal upper critical dimension
$d<d_{\mathrm{uc}}=4$. Second, their study is more difficult due to
the dimensional reduction phenomenon, which renders naive
zero-temperature perturbation theory trivial, hence useless. The way
out involves first parameterizing the set of relevant operators into a
function, $\Delta(u)$, the second cumulant of the random pinning
force. It was shown in Refs.~\cite{nstl92,narayan-fisher93} that the
corresponding functional renormalization group (FRG) provides an
adequate description of the depinning transition if one considers the
non-analytic renormalized function $\Delta(u)$.  It was shown only
recently that the FRG can be unambiguously extended to higher loop
order and that the underlying non-analytic field theory is
renormalizable \cite{chauve01,ledoussal02}.  The FRG equation for
$\Delta(u)$ has two main non-trivial stable fixed points which
describe periodic and interface universality classes. Both of them
exhibit a cusp singularity of the form $\Delta^*(u)-\Delta^*(0)\sim
|u|$ at small $u$. This cusp accounts for the existence of the
critical threshold force $f_c\sim \Delta^{*\prime}(0^+)$. The
corresponding critical exponents have been computed to second order in
$\varepsilon=4-d$ \cite{chauve01,ledoussal02}.

Despite this progress many open questions remain. Among them is the
problem of sample-to-sample fluctuations, i.e.\ the probability
distributions of a given observable and their relation to extreme
value statistics.  These were studied mostly for static
quantities. The distribution of the energy of pinned manifolds was
analyzed in Ref.~\cite{alava03,fedorenko03}. The distribution of the
mean squared width of an interface at the depinning transition was
calculated using a Gaussian approximation for the displacement field,
yielding a result quite close to numerics \cite{rosso03}. It was shown
how systematic corrections can be computed in the field theory of
depinning within an $\epsilon=4-d$ expansion \cite{ledoussal03}. One
expects sample-to-sample fluctuations to play an important role in the
dynamics too, leading to a broad distribution of time scales. The
divergence of the typical energy barrier with scale, responsible for
the ultraslow creep motion, is predicted by phenomenological arguments
\cite{vortex}. A numerical study in Ref.~\cite{drossel} of the
distribution of barriers confirms that the typical barrier scales as
the energy minima, as predicted by 1-loop FRG studies
\cite{creep}. The more difficult question of predicting the
distribution of energy barriers was addressed in
Ref.~\cite{ledoussalbalents} using the FRG, and in
Ref.~\cite{vinokur-96} using extreme value statistics.

An important and debated question is to characterize the finite-size
fluctuations of the critical force and whether they obey finite-size
scaling (FSS). Similar questions were investigated recently in the
context of heteropolymer unbinding transitions (the role of critical
force being played there by critical temperature) where violation of
FSS was found \cite{monthus05}. In the depinning problem, one
difficulty is to define properly the critical force and its
fluctuations in the limit of large interface (internal) size $L$.  A
recent and efficient algorithm \cite{rosso2001} allows to obtain
exactly the critical force $f_c$ of an interface in a periodic medium
of period $M$ (i.e.\ a cylinder), as well as the so-called critical
configuration. The latter is defined as the last blocking
configuration as $f$ is increased up to $f_c$, which also defines
$f_c=f_c(L,M)$. One can refer to this definition as an extremal
configuration in a given sample. The finite-size sample-to-sample
distribution of $f_c(L,M)$ was computed numerically \cite{bolech04}
and found to depend on the aspect ratio $k=M/L^\zeta$ of the
cylinder. This should be expected since for large $k$ one recovers a
zero-dimensional problem and the interface will be blocked by rare
disorder configurations, hence dominated by extremal
statistics. However this results seems to depend on the precise
definition and one may ask whether a more fundamental definition
exists, with no need to specify a value for $k$. An alternative
approach is to define the observables at the depinning transition as
the time average in the moving state at fixed velocity $v$, in the
limit $v \to 0^+$. This definition, to which we refer as the
quasi-static depinning limit, is usually associated to the FRG
approach of the depinning transition. Observables calculated in this
approach must, a priori, be distinguished from those computed in the
critical configuration. Since the time average is usually performed in
a steady state, to avoid dominance by history dependence, it also
requires specifying boundary conditions. It is widely believed that
both approaches give the same result, at least for ($N=1$ component)
interfaces, since in the limit of infinite systems ($L \to \infty$)
all quasi-static configurations should have the same statistical
properties and the critical force $f_c$ should be self-averaging.
However it is less clear how these approaches compare when applied to
finite-size fluctuations where the dispersion in local pinning forces
becomes important.

In the present paper we study the distribution of the threshold forces
by means of the functional renormalization group. Within the field
theory we propose two definitions of the critical force $f_c(L)$ in
finite size $L$ and show that they are identical to one loop in the
renormalized theory.  We compute the cumulants of $f_c(L)$ and extract
the distribution which is found to be universal, up to a shift in $f$
(the critcal force $f_{c}$) and one scale-parameter (the width of the
distribution). All results are valid within the $\epsilon=4-d$
expansion and extrapolations to low dimension are discussed. The
critical force studied here is defined from a fixed center of mass
ensemble. As we point out it can be, in principle, obtained in
numerics. Since it does not refer to any transverse size $M$, it is
more fundamental than the one used in the numerical studies on a
cylinder.  We discuss how the latter one can in principle be
recovered.

The paper is organized as follows. Section \ref{sec1} introduces the
model and the FRG treatment of the depinning transition. In Section
\ref{sec2} we compute the bare distribution of threshold forces to
one-loop order using an improved perturbation theory, and renormalize
it for the case of an elastic interface. In Section \ref{sec3} we
discuss the renormalization for periodic systems.  In Section
\ref{sec4} we discuss the relation between the distribution of
critical forces, in the quasi-static limit, and in the critical
configuration.

\section{Model and FRG description}
\label{sec1}

Let us consider the motion of a one-component elastic manifold with
short-range elasticity. The configuration of the manifold can be
described by a scalar displacement field $u_{xt}$, where $x$ denotes
the $d$ dimensional internal coordinate of the manifold. We study the
over-damped dynamics of a manifold in the disordered medium which
obeys the following equation of motion
\begin{equation}
\eta\partial_t u_{x t} = c\nabla^2 u_{x t} + F(x,u_{x t})+f,
\end{equation}
where $\eta$ is the bare friction and $c$ the elasticity. The
quenched random force $F(x,u)$ can be chosen Gaussian with zero mean
and variance
\begin{equation}
\overline{F(x,u)F(x',u')}=\Delta(u-u')\delta^d(x-x'). \label{FF}
\end{equation}
For periodic systems the function $\Delta(u)$ is periodic, while for
interfaces it decays exponentially for large $u$. In the latter case,
in contrast to the statics, at depinning both
random bond (RB) and random field (RF) microscopic disorder renormalize to the
same fixed point, which has RF characteristics, so that we
can restrict ourselves to the latter case. $a$ is the width of the
function  $\Delta(u)$. To make the problem well-defined we imply an
UV cutoff  at scale $\Lambda^{-1}$. We consider a finite system
of size $L$ with periodic boundary conditions. The size $L$ serves
as an IR cutoff i.e.\  it plays the role of the mass in the
corresponding field theory. One can easily see that due to the tilt
symmetry the elastic constant  remains uncorrected to all orders so
that we are free to fix $c=1$.

Below, starting in Section \ref{sec2} we will find it convenient to
work in the comoving frame. To that end we  shift $u_{xt}\to v t
+ u_{xt}$, s.t.\ $\overline{\langle u_{xt}\rangle}=0$ and
$f \to f- \eta v$, where $v=L^{-d}\overline{\langle\int_x\partial_t
u_{x t}\rangle}$ is the velocity of the center of mass. Here the
angular brackets stand for the average over different initial
configurations (since we are studying zero temperature dynamics) and
the overline denotes the average over disorder distribution. We will
assume that a steady state attractor has been reached, hence that
averages depend only on time differences and not on a specific
choice of initial conditions.

To study the dynamics of an elastic manifold efficiently, we
use the formalism of generating functionals. Introducing the
response field $\hat{u}_{xt}$ one can compute the average of the
observable $A[u_{x,t}]$ over dynamic trajectories with different
initial conditions for a particular disorder configuration as
follows
\begin{equation}
\langle A[u_{x,t}] \rangle = \int \mathcal{D}[u] \mathcal{D}[\hat{u}] A[u_{x,t}]
e^{-S_{F}[u,\hat{u}]}\ .
\end{equation}
$S_{F}$ is the action  for a particular realization of the disorder
(a particular sample). To compute the average of the observables which
explicitly depends on the random force at the position of the manifold we introduce
the source $J_{x t}$ for the random force $F$ so that the corresponding action reads
\begin{eqnarray}
&& \!\!\!\!\!\!\!\!\!\!\!\!
  S_{F}[u,\hat{u}]  = \int_{x t} i\hat{u}_{x t}(\eta\partial_t - \nabla^2 )u_{x t}
  \nonumber \\
&& \ \
- \int_{x t} i\hat{u}_{x t} \{ F(x, u_{x t}) + f_{x t}\}-\int_{x t} J_{x t}  F(x, u_{x t} ).
\end{eqnarray}
After averaging over the disorder distribution, any observable which depends on
the displacement field and the random force at the position of the manifold can be
computed as follows
\begin{eqnarray}
&&\overline{\langle A[u_{x,t}] F(x_1,u_{x_1,t_1})...F(x_n,u_{x_n,t_n}) \rangle} \nonumber \\
&& = \left.
\frac{\delta}{\delta J_{x_1,t_1}...\delta J_{x_n,t_n}}
\int \mathcal{D}[u] \mathcal{D}[\hat{u}] A[u_{x,t}] e^{-S[u,\hat{u}]}
\right|_{J=0} , \nonumber
\end{eqnarray}
$S[u,\hat{u}]$ is the effective action, which can be split into two parts:
the free part $S_0$ being quadratic in fields  and the interaction part
$S_{\mathrm{int}}$ containing all non-linear terms
\begin{eqnarray}
&&S_{0}=\int_{x t} i\hat{u}_{x t}(\eta\partial_t - \nabla^2 )u_{x t}-
\int_{x t} i\hat{u}_{x t} f_{x t}, \nonumber \\
&&S_{\mathrm{int}}=-\frac12\int_{x t t'}(i\hat{u}_{x t}+J_{x t})
\Delta({u}_{x t}-{u}_{x t'})(i\hat{u}_{x t'}+J_{x t'}). \nonumber
\end{eqnarray}
Setting $J_{xt}=0$ we recover the action used in
Refs.~\cite{chauve01,ledoussal02}. The quadratic part $S_0$ gives
the free response
\begin{eqnarray} \label{propagator}
 \langle u_{q,t}\ i\hat{u}_{-q,0}  \rangle = R_{q,t}=\frac{\Theta(t)}{\eta} e^{-q^2t/\eta},
\end{eqnarray}
while the free correlation function is $C_{q,t}=\langle u_{q,t}\
{u}_{-q,0}  \rangle=0$ at zero temperature. The splitted
diagrammatics for the perturbation theory in disorder $\Delta$ was
developed in Refs.~\cite{chauve01,ledoussal02}. It is known that
naive perturbation theory, obtained by taking for $\Delta$ an
analytic function exhibits the property of dimensional reduction
and fails to describe the physics, giving for example an incorrect
roughness exponent. The physical reason for this is the existence
of a large number of metastable  states.

Let us briefly sketch the FRG analysis of the system under
consideration. Power counting shows that the whole function
$\Delta(u)$ becomes relevant below $d_{\mathrm{uc}}=4$ and thus one
has to renormalize the whole function. To extract the scaling behavior
one has to study the flow of the renormalized function $\Delta$ under
changing the IR cutoff towards infinity. Various choices for the IR
cutoff were discussed in Refs.\cite{chauve01,ledoussal02}. A
convenient choice is to add a small mass $m$, so that the scaling
behavior can be extracted from the effective action
$\Gamma[u,\hat{u}]$ of the theory as $m$ decreases to zero. To study
the finite-size distribution of threshold forces, we use, as in
\cite{ledoussal03}, the system size $L$ as the natural IR cutoff.
Then any integral over momentum $q$ has to be replaced by the sum
according to the rule $\int_q \to L^{-d} \sum_{q}$, where the sum runs
over all $q=2\pi k/L$, $k \in \mathbb{Z}^d,k\neq 0$. Exclusion of the
zero mode means that we are working in an ensemble of fixed center of
mass, a point further discussed in Section \ref{sec4}.

Let us define the rescaled disorder as
\begin{equation} \label{equ1}
\Delta(u)=\frac{1}{\varepsilon \tilde{I}_1} L^{2\zeta-\varepsilon}\tilde{\Delta}(uL^{-\zeta})\ ,
\end{equation}
where $I_1=L^{\varepsilon}\tilde{I}_1=\int_{{q}}|{q}|^{-2}$ is the
one-loop integral. It was shown in
Refs.~\cite{nstl92,narayan-fisher93} that the FRG equation, i.e.\ the
flow equation for the running disorder correlator can be written to
one-loop order as
\begin{eqnarray}
\left.L\partial_L \tilde{\Delta}(u)\right|_0&=&(\varepsilon-2\zeta)\tilde{\Delta}(u)+\zeta u\tilde{\Delta}'(u)
\nonumber\\
&-&\frac12 \left[\left(\tilde{\Delta}(u)-\tilde{\Delta}(0)\right)^2 \right]'', \label{FRG}
\end{eqnarray}
the two loop flow equation being obtained in
Refs.~\cite{chauve01,ledoussal02}.  ``0'' means a derivative at fixed
bare quantities. The flow of the correlator is such that $\Delta (u)$
acquires a cusp at the origin $u=0$ at the Larkin scale $L_{c}\simeq
\left( c ^{2}a^{2}/\Delta (0)\right) ^{1/\varepsilon }$.  Beyond the
Larkin scale ($L>L_{c}$) the renormalized correlator is singular and
perturbation theory breaks down. Nevertheless, the flow tends to a
non-trivial fixed-point (FP) solution $\tilde{\Delta}^*(u)$ with a new
value for the roughness exponent which controls the large-scale
behavior. There are two FPs which describe interfaces and periodic
systems correspondingly. The former FP has
$\zeta=\varepsilon/3+\mathcal{O}(\varepsilon^2)$, while the latter one
has $\zeta=0$ due to periodicity. The renormalization of the mobility
gives the value of the dynamic exponent
\begin{equation}
  z = 2 + \left. L \frac{d   }{d L} \eta_R \right|_{0} = 2-\tilde{\Delta}''(0)
   +\mathcal{O}(\tilde{\Delta}^2),
\end{equation}
where $\eta_R$ is the renormalized mobility.
Other critical exponents can be computed using the scaling relations
\begin{equation}
\nu= \frac1{2-\zeta}=\frac{\beta}{z-\zeta}.
\end{equation}
To renormalize the theory, one needs an additional counter-term for
the excess force $f-\eta v$, which comes with an UV divergence $\sim
\Lambda^2$. This term is analogous to the critical temperature shift
in the $\varphi^4$ theory, and gives the critical threshold force
$f_c^*$. It is zero in the bare theory. We expect that in the limit of
an infinite system $L \to \infty$, the critical force becomes sample
independent if there is the same distribution of disorder in each
sample and thus $\lim_{L\to \infty} P_L(f)=\delta(f-f_c^*)$. However
the situation is different for finite systems. According to the
general theorem for random systems \cite{chayes86} there exists a
finite-size scaling correlation length $\xi_{\mathrm{FS}}$ which
characterizes the distribution of the observables in an ensemble of
samples and which in principle has to be distinguished from the
intrinsic correlation length $\xi$ which enters into correlation
functions. Approaching the critical point, the finite-size correlation
length diverges similar to the intrinsic correlation length as
$\xi_{\mathrm{FS}}\sim |f-f_c|^{-\nu_{\mathrm{FS}}} $. In general
$\nu_{\mathrm{FS}}$ is different from $\nu$, and satisfied the
inequality
\begin{equation}\label{nufs}
\nu_{\mathrm{FS}}\ge 2/(d+\zeta)\ ,
\end{equation}
where $d+\zeta$ is the effective dimension of the disordered system
considered.  Thus for an ensemble of samples of linear size $L$ the
width of the distribution of critical forces is characterized by a
scale $\xi_{\mathrm{FS}}=L$ and reads
\begin{equation}
\overline{\langle(f_c(L)-f_c^*)^2\rangle}\sim L^{-2/\nu_{\mathrm{FS}}}
\end{equation}
For periodic systems $\zeta=0$ and $\nu=1/2$ so that
$\nu_{\mathrm{FS}}\neq \nu$ for $d\leq 4$. For interfaces it was
proposed ~\cite{narayan-fisher93}
that $\nu_{\mathrm{FS}}=\nu$. While $\nu_{\mathrm{FS}}$
satisfies (\ref{nufs}) with the equal-sign in 1-loop order, at 2-loop
order the inequality becomes strict.
As discussed in
Refs.~\cite{narayan-fisher93,ledoussal02} this difference is closely
related to the stability of the FP which controls the scaling
behavior.

\section{Distribution of threshold forces}
\label{sec2}
\subsection{Perturbation theory}
Let us show how the critical force distribution can be computed within
``improved'' perturbation theory. ``Improved'' means that we assume
the the disorder correlator $\Delta(u)$ to be non-analytic, since for
analytic disorder  perturbation theory gives a zero-threshold
force.  Then using  FRG we  renormalize our result to 1-loop
order.  Effectively, this is a  summation of an infinite subset of
diagrams.  We define the distribution of  threshold force as
follows
\begin{equation}
P_L(f_c)=\overline{\left\langle\delta\left(f_c+L^{-d}\int_x
F(x,vt+u_{xt})\right)\right\rangle}\ . \label{distr-def}
\end{equation}
From now on we work in the comoving frame and the average is performed
with the action $S$ in the quasi-static limit $v \to 0^+ $, as is
usually done in the FRG theory of the depinning transition.

\begin{figure}[tbp]
\includegraphics[clip,width=2.2in]{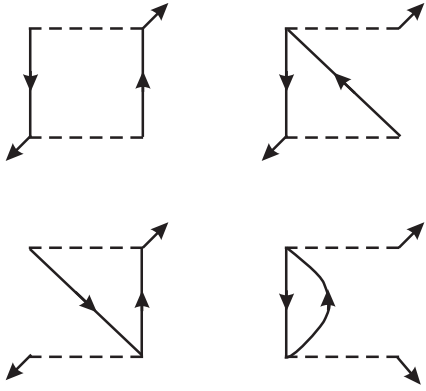}
\caption{Diagrams $D_i$, ($i=1,...,4$)
contributing to the second cumulant of the threshold force.
We have adopted the splitted diagrammatics used in Ref.~\cite{ledoussal02}:
arrowed line indicates a response propagator (\ref{propagator});
dashed line - the splitted vertex
$i\hat{u}_{x t}i\hat{u}_{x t'} \Delta({u}_{x t}-{u}_{x t'})$.
The corresponding expressions are given by Eq.~(\ref{eq4}).}
\label{fig0}
\end{figure}

Let us introduce the corresponding characteristic function
\begin{equation}
\hat{P}_L(\lambda)=\overline{\langle e^{-i\lambda f_c(L)}\rangle} =
\int df_c e^{- i \lambda f_c} P_L(f_c)
\end{equation}
which can be expressed through the cumulants
$\overline{\langle f_c(L)^n\rangle}_{c}$ as follows
\begin{equation}
\hat{P}_L(\lambda)=\exp\left( \sum\limits_{n=1}^{\infty}
\frac{(-i\lambda)^n}{n!} \overline{\langle f_c(L)^n\rangle}_{c} \right).
\end{equation}
The computation of the first cumulant to one loop is trivial.  The
random force that the interface  actually feels in the point $x$ is
given by
\begin{eqnarray}
\!\!\!\!\!\! &&  \overline{\langle F(x,u_{x,t}+vt)\rangle}   =
\left\langle \int_{t_1} \Delta(u_{xt}-u_{xt_1}+v(t-t_1))i\hat{u}_{x
t_1}\right \rangle
 \nonumber \\
&& \ \ \ \ \ \  =  \int_{t_1} \Delta'(v(t-t_1))[R_{x=0,t-t_1}-R_{x=0,t=0}].
\label{mean-value}
\end{eqnarray}

\noindent We will adopt  Ito's prescription in which
$R_{x,t=0}=0$. Note that this corresponds to the definition  $\Theta(0)=0$.
Taking the quasi-static limit $v \to 0^+$, we obtain the well-known
expression for the average critical force
\begin{equation}
f_c^*=-\Delta'(0^+) \int\limits_{0}^{\infty} dt R_{x=0,t}
=- \int_q \frac{\Delta'(0^+)}{q^2}. \label{f-mean}
\end{equation}
Note that the critical force ({\ref{f-mean}}) diverges at large
momentum as $\Lambda^{d-2}$ and therefore is not universal, i.e.\ it
depends on microscopic parameters. This is analogous to to the shift
of the critical temperature in standard critical phenomena, caused by
fluctuations.  As we know this is shift is also non-universal. However,
we expect that the distribution of critical forces for a finite system
around the average value are universal, once the distribution is
properly normalized.  The computation of the the $n$-th cumulant is
more tricky. Before considering the general case let us show how this
works for the second cumulant. Using the generating functional, we can
write down the formal expression for the effective force-force
correlator
\begin{eqnarray}
&& \!\!\!\!\!\!\!\! \overline{\langle
F(x_1,u_{x_1,t}+vt)F(x_2,u_{x_2,t}+vt)\rangle}  =
\Delta(0)\delta^d(x_1-x_2) \nonumber \\
&& \ \ \ \ \ \ +  \left\langle
\int_{t_1} \Delta(u_{x_1t}-u_{x_1t_1}+v(t-t_1))i\hat{u}_{x_1 t_1} \right. \nonumber \\
&& \ \ \ \ \ \  \left. \times \int_{t_2} \Delta(u_{x_2t}-u_{x_2t_2}
+v(t-t_2))i\hat{u}_{x_2 t_2}
\right \rangle. \ \ \label{second}
\end{eqnarray}
The first term on the r.h.s.\ of Eq.~(\ref{second}) is the bare
disorder distribution. It is given by Eq.~(\ref{FF}) and is a pure
Gaussian distribution with zero mean. However the moving manifold
explores a different distribution that is an effective distribution
which one can observe ``sitting'' on the moving interface. The second
term on the r.h.s.\ of Eq.~(\ref{second}) as well as the mean value
(\ref{mean-value}) are the deviation of the effective distribution
from the bare one.  Only connected diagrams contribute to the second
cumulant.  Integrating the second term in Eq.~(\ref{second}) over
fields with the weight $e^{-s}$ we obtain the four connected diagrams
shown in Figure \ref{fig0}.  The corresponding expressions can be
rewritten as follows
\begin{eqnarray}
  && \mbox{} \hspace{-7mm}\!\!\!\ \int_{t_1t_2}
\Delta'(v(t-t_1))[R_{x_2-x_1,t-t_1}-R_{x_2-x_1,t_2-t_1}]\nonumber \\
&& \ \ \Delta'(v(t-t_2))[R_{x_1-x_2,t-t_2}-R_{x_1-x_2,t_1-t_2}].
\end{eqnarray}
To compute the contribution to the variance of the critical force, we
have to integrate over $x_1$ and $x_2$ and then multiply by $L^{-2d}$.
This computation is more convenient in Fourier representation:
\begin{eqnarray}
&& D_1+D_2+D_3+D_4= L^{-d}[\Delta'(0^+)]^2 \nonumber \\
&& \times \int_{q t_1 t_2} [R_{q,t-t_1}-R_{q,t_2-t_1}]
[R_{q,t-t_2}-R_{q,t_1-t_2}]. \ \   \label{eq4}
\end{eqnarray}
Due to causality we have $D_4=0$. The other diagrams read
\begin{eqnarray}
D_1=-D_2=-D_3= L^{-d} \int_q \frac{[\Delta'(0^+)]^2}{(q^2)^2}\ .
\end{eqnarray}
Summing all contributions we obtain
\begin{eqnarray}
\overline{\langle f_c(L)^2\rangle}_c =L^{-d}\Delta(0) -
L^{-d} \int_q \frac{[\Delta'(0^+)]^2}{(q^2)^2}. \label{f-2-bare}
\end{eqnarray}
We have also derived Eq.~(\ref{f-2-bare}) by using direct perturbation
theory instead of the generating functional.

The above calculation can be generalized to arbitrary $n$.
It can be  simplified significantly if one takes
into account that all intermediate times $t_i$ must be smaller than
the observation time $t$:  $t_i<t$ ($i=1,..,n$).
For the $n$-th cumulant ($n>2$) we have
\begin{widetext}
\begin{eqnarray}
 \overline{\langle f_c(L)^n\rangle}_c &=& (-1)^n (n-1)! L^{-d(n-1)} \int_{q}
  [\Delta'(0^+)]^n \nonumber \\
 &\times& \int_{t_1...t_n}
[R_{q,t-t_1}-R_{q,t_2-t_1}]\
[R_{q,t-t_2}-R_{q,t_3-t_2}] ...
[R_{q,t-t_{n-1}}-R_{q,t_n-t_{n-1}}]\
[R_{q,t-t_n}-R_{q,t_1-t_n}]. \label{Fn-c1}
\end{eqnarray}
\end{widetext}
Here the factor $(n-1)!$ results from different contractions of
$u_{x_it}-u_{x_it_i}$ and $\hat{u}_{x_jt_j}$ ($i,j=1,\dotsc, n$) that
form a closed loop. Expanding the integrand in Eq.~(\ref{Fn-c1}) we
find that all terms gives the same contribution up to a factor of $\pm
1$, except for the term composed only from the second response
function in each bracket. This term gives a closed loop of
response-functions, which is zero by causality. Using the identity
\begin{equation}
 \sum\limits_{i=0}^{n-1}(-1)^i C_n^i=(-1)^{n+1},
\end{equation}
where $C^i_n$ is a binomial coefficient, we can simplify  Eq.~(\ref{Fn-c1}) to
\begin{equation} \label{f-n-bare}
\overline{\langle f_c(L)^n\rangle}_c = -
(n-1)! L^{-d(n-1)} \int_{q} \frac{[\Delta'(0^+)]^n}{q^{2n}}.
\end{equation}
We are now in a position to construct the characteristic function
\begin{eqnarray}
\ln \hat{P}(\lambda)&=& -\frac12\Delta(0)L^{-d}\lambda^2 \nonumber \\
&-& L^d \int_q \sum\limits_{n=1}^{\infty}
\frac{(-1)^n}{n} \left(  \frac{\Delta'(0^+)}{q^2} L^{-d} i\lambda \right)^n.
\label{res0}
\end{eqnarray}
The latter is nothing but the Taylor series of the logarithm, which
allows to rewrite Eq.~(\ref{res0}) as
\begin{eqnarray}
\hat{P}(\lambda)&=&\exp\left[-\frac12\Delta(0)L^{-d}\lambda^2 \right.\nonumber \\
&+& \left. L^d \int_q
\ln \left(1 - \frac{|\Delta'(0^+)|}{q^2} L^{-d} i\lambda \right) \right],
\label{res1}
\end{eqnarray}
where we have taken into account that $\Delta'(0^+)<0$.

As follows from the above computation, the distribution of the
critical force can be related to the effective action
$\Gamma[u,\hat{u}]$ which is a generating functional for one-particle
irreducible (1PI) vertex functions $\Gamma^{(E,S)}_{\hat{u}\dotsb
\hat{u};u\dotsb u}$ with $S$ external fields $u$ and $E$ external
fields $\hat{u}$
\begin{eqnarray}
&& \!\!\!\!\!\! \Gamma^{(E,S)}_{\hat{u}..\hat{u};u..u}(\{\hat{q}_i,\hat{\omega}_i\},
 \{q_j,\omega_j\})  \nonumber \\
&& =\left. \prod\limits_{i=1}^{S} \frac{\delta}{\delta u_{q_i,\omega_i}}
\prod\limits_{j=1}^{E} \frac{\delta}{\delta \hat{u}_{\hat{q}_j,\hat{\omega}_j}}
\Gamma[u,\hat{u}] \right |_{u=\hat{u}=0}\ .
\label{generating}
\end{eqnarray}
Indeed as already seen from the bare action the average threshold
force can be expressed as vertex function
$\Gamma^{(1,0)}_{\hat{u}}(q=0,\omega=0)$ in the quasi-static limit $v
\to 0^+$.  Analogously the higher-order cumulants can be identified as
the higher-order vertex-functions according to
\begin{equation} \label{f-n-vertex}
\overline{\langle f_c(L)^n\rangle}_c = L^{-(n-1) d}
\ \Gamma^{(n,0)}_{\hat{u}\dotsb \hat{u}}(\{q_i=0,\omega_i=0\}).
\end{equation}
The general properties of vertex functions (\ref{f-n-vertex}) for even
$n$ was discussed in Ref.~\cite{ledoussal03}. In particular it was
noted that loop diagrams that contribute to the vertex
$\Gamma^{(2n,0)}$ precisely cancel each over so that the result is
given by minus the missing contribution from acausal loops.
It is easy to verify by direct inspection of the Feynman diagrams that
definitions (\ref{distr-def}) and (\ref{f-n-vertex}) give the the same result in
the one-loop approximation,
but the question of their equivalence to all orders is open.

\begin{figure}[tbp]
\includegraphics[clip,width=2.8in]{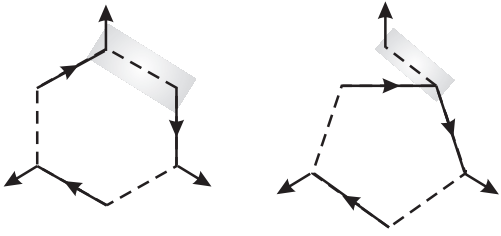}
\caption{Example of diagrams contributing to the third cumulant of the
critical force. To renormalize these one-loop diagrams to the lowest
order we replace the disorder lines by vertices
$\Gamma^{(+)}_{\hat{u}\hat{u}u}$ (highlighted on the left diagram) and
$\Gamma^{(-)}_{\hat{u}\hat{u}u}$ (highlighted on the right diagram)
which are defined in Eqs.~(\ref{vertex4}) and (\ref{vertex5}) and
depicted in details in Figure~\ref{fig1}.}  \label{fig-g3}
\end{figure}

\begin{figure}[tbp]
\includegraphics[clip,width=2.8in]{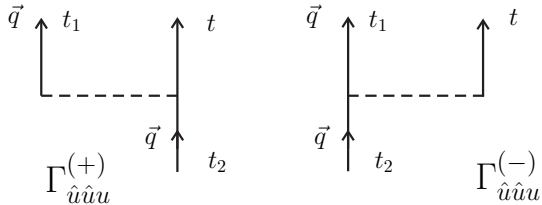}
\caption{Vertex function
$\Gamma^{(2,1)}_{\hat{u}\hat{u}u}(t,t_1,t_2;q,-q)$ at the tree
level. The functions are distinguished by whether the line entering at
$t_{2}$ and outgoing at $t_{1}$ is ``closed''
$\Gamma^{(-)}_{\hat{u}\hat{u}u}$, or ``open''
$\Gamma^{(+)}_{\hat{u}\hat{u}u}$. In the first case the time-ordering
along the loop is continuous: $t_1<t_2$, while in the second case it
is interrupted, i.e.\ there are no restrictions on $t_{1}$ and $t_{2}$
(see also Figure~\ref{fig-g3}).} \label{fig1}
\end{figure}

\subsection{Renormalization}
In this section we focus on the interface
problem, periodic systems being considered in the next section. The
distribution of the critical forces in (\ref{res1}) has been obtained
from the ``improved'' perturbation theory, and thus, it cannot be
reproduced within the Larkin type models in which all observables
depend only on $\Delta(0)$. However in the bare theory the disorder
correlator is an analytic function so that to make the calculation
consistent we have to first renormalize our theory. To that end we
replace the bare correlator by the renormalized one. This can be
viewed as a partial summation of an infinite series of diagrams. If we
want the distribution of $f_c$ strictly to lowest order in
$\epsilon=4-d$ then the work is essentially done. However, we will
demand a bit more and take an additional effect into account:
replacing the bare correlator by the running one in a particular
diagram we have to be careful because the scale dependence acquired by
the correlator may be in form of either dependence on a mass (as in
quantities which do not contain integration over momentum) or
dependence on the loop momentum (which has to be integrated out), or a
combination of both. The final result presented here will thus be
exact to lowest order in $\epsilon$ and in addition will contain some
effects beyond that order (although a full fledge two loop calculation
is not attempted here). This will allow us to discuss in the next
section some extrapolations  to low dimension.

Let us start from the renormalization of the first cumulant, i.e.\
the average critical force (\ref{f-mean}).  We remind the reader
that the average critical force is a non-universal quantity and
afterwards we will subtract it and consider the shifted distribution
which is a universal function. According to Eq.~(\ref{equ1}) the
renormalized disorder correlator acquires in the vicinity of the
fixed-point a scale dependence. Integration over scales beyond the
Larkin scale yields (see Ref \cite{creep,movglass} for details):
\begin{eqnarray}
\overline{\langle f_c(L)\rangle}_c \approx  - {\tilde{\Delta}^{*\prime}  (0^+)}
   \frac {\Lambda^{2-\zeta}}{2-\zeta}
\end{eqnarray}
where in this formula the UV cutoff $\Lambda$ is meant to be the
minimal length of pinned segments of the manifold, i.e.\ , the Larkin
length $\Lambda \sim L_c^{-1}$.

We now consider the renormalization of the second cumulant
(variance). The corresponding bare expression can be expressed through
the 2-point vertex function as follows
\begin{eqnarray}
 \overline{\langle f_c(L)^2 \rangle}_c  &=& L^{-d}\left[\Delta(0)-
\int_q  \frac{\Delta'(0^+)^2}{q^4 } \right] + O(\Delta^3) \nonumber \\
&=& L^{-d}\ \Gamma^{(2,0)}_{\hat{u}\hat{u}}(\omega=0;q=0).
\label{ren1}
\end{eqnarray}
As it was shown in Ref.~\cite{ledoussal02,ledoussal03}, the 2-point
vertex function does not depend on times and scales,
\begin{eqnarray}
\Gamma^{(2)}_{\hat{u}\hat{u}}(q)= L^{2\zeta-\varepsilon}\frac1{\tilde{I}_1
 \varepsilon} \tilde{\Delta}^*(0)F_2\left({q}{L}\right), \label{ren2}
\end{eqnarray}
with $F_2(z)=B z^{\varepsilon -2\zeta} + O(\ln z/z^2)$ for large $z$
and $F_2(0)=1$.  Note the the constant $B$ depends on the IR cutoff
scheme \cite{ledoussal03}.  Combining Eqs.~(\ref{ren1}) and
(\ref{ren2}) we obtain
\begin{eqnarray}
\overline{\langle f_c(L)^2 \rangle}_c  = L^{2\zeta-4}\frac1{\tilde{I}_1
 \varepsilon} \tilde{\Delta}^*(0).
\end{eqnarray}
We note that this is consistent with the finite-size scaling prediction:
\begin{eqnarray}
\overline{\langle f_c(L)^2 \rangle}_c  \sim L^{-2/\nu}
\end{eqnarray}
using $\nu=1/(2-\zeta)$. As we will show below, the full (shifted)
distribution is also consistent with this scaling.

To proceed further, let us first consider some typical diagrams
contributing to the third cumulant of the critical force, which are
shown in Figure~\ref{fig-g3}. To renormalize them at lowest order, we
have replaced the disorder lines by the three-point vertices defined
as follows
\begin{eqnarray}
\Gamma^{(2,1)}_{\hat{u}\hat{u}u}(t,t_1,t_2;q_1,q_2)&=&
\Gamma_{\hat{u}\hat{u}u}^{(+)}(t, t_1,t_2;q_1,q_2) \nonumber \\
&+&
\Gamma_{\hat{u}\hat{u}u}^{(-)}(t, t_1,t_2;q_1,q_2).\label{vertex3}
\end{eqnarray}
At tree level, the vertex function (\ref{vertex3}) can be expressed by
diagrams shown in Figure \ref{fig1} and the corresponding expressions
read
\begin{eqnarray}
\Gamma_{\hat{u}\hat{u}u}^{(+)}(t,t_1,t_2 ;q_1,q_2)&=&
\Delta'(0^+) \mathrm{sign}(t-t_1) \delta (t-t_2),\ \  \ \  \label{vertex4} \\
\Gamma_{\hat{u}\hat{u}u}^{(-)}(t,t_1,t_2 ;q_1,q_2)&=&
\Delta'(0^+) \mathrm{sign}(t_1-t) \delta (t_1-t_2). \ \ \ \ \label{vertex5}
\end{eqnarray}
Then the summation of diagrams contributing to the the $n$-th cumulant
with $n>2$ can be carried out along the lines used for the bare
cumulant and gives
\begin{widetext}
\begin{eqnarray}
&& \mbox{} \hspace{-10mm}
 \overline{\langle f_c(L)^n\rangle}_c = (-1)^n (n-1)! L^{-d(n-1)} \int_{q}
 \int_{t_1...t_n}\int_{\tau_1...\tau_n}
\left[\Gamma_{\hat{u}\hat{u}u}^{(2,1)}(t,t_2,\tau_1 ;q,-q) R_{q,\tau_1-t_1}\right]\
\times \nonumber \\
&&
\left[\Gamma_{\hat{u}\hat{u}u}^{(2,1)}(t,t_3,\tau_2 ;q,-q) R_{q,\tau_2-t_2}\right]\
...
\left[\Gamma_{\hat{u}\hat{u}u}^{(2,1)}(t,t_n,\tau_{n-1} ;q,-q) R_{q,\tau_{n-1}-t_{n-1}}\right]\
\left[\Gamma_{\hat{u}\hat{u}u}^{(2,1)}(t,t_1,\tau_n ;q,-q) R_{q,\tau_n-t_n}\right].
\label{f-n-cumulant}
\end{eqnarray}
\end{widetext}
Substituting the tree-level expressions
(\ref{vertex3})-(\ref{vertex5}) to Eq.~(\ref{f-n-cumulant}) we recover
the bare cumulant (\ref{Fn-c1}).

In Appendix~\ref{appendix} we compute the vertex function
$\Gamma_{\hat{u}\hat{u}u}^{(+)}$ to one-loop order and obtain its
large-$q$ asymptotics which reads
\begin{eqnarray}
\int_{t_2}\Gamma_{\hat{u}\hat{u}u}^{(+)}(t,t_1,t_2 ;q,-q) &=& A
L^{\zeta-\varepsilon} \left({q}L\right)^{\psi}, \label{Gamma3}
\end{eqnarray}
where times $t_1$ and $t$ are taken infinitely apart (hence it is a
quasi-static quantity). The amplitude $A$ and the exponent $\psi$ are
given by
\begin{eqnarray}
   A&=&\frac{1}{\varepsilon\tilde{I}_1}\tilde{\Delta}'^*(0^+)
                             (1+\mathcal{O}(\varepsilon)), \label{AA} \\
  \psi &=& \frac49\varepsilon + \mathcal{O}(\varepsilon^2),
\label{psi}
\end{eqnarray}
and we argue that $\psi$ is a new exponent (see
Appendix~\ref{appendix}). Taking into account this momentum dependence
in Eq.~(\ref{f-n-cumulant}) should result in an improved
renormalization scheme (compared to simply replacing the bare
quantities by $q$-independent but scale-dependent renormalized
parameters) with a different form for the $q$ summations appearing
below.

After substituting Eq.~(\ref{Gamma3}) in Eq.~(\ref{vertex3}), the
integration over times can be performed in the same way as for the
bare cumulant and gives for $n>2$\footnote{Note that not taking into account
$q$ dependence in $\Gamma_{\hat{u}\hat{u}u}^{(+)}(t,t_1,t_2 ;q,-q)$
amounts to choose $\psi=0$ while the ansatz used in the calculation of
the width distribution \cite{ledoussal03} , i.e.\ replacing $q^{-4}
\to q^{-d-2 \zeta}$ would amount to choose $\psi=\epsilon/6$. }:
\begin{eqnarray}
\overline{\langle f_c(L)^n\rangle}_c &=& - (n-1)!\, A^n L^{n(\zeta-2)} \nonumber \\
&  & \times L^d \int_{q} \frac1{(qL)^{n(2-\psi)}}, \label{f-n-ren} 
\end{eqnarray}
Note that $A\sim \tilde{\Delta}^{*\prime}(0^+)<0$. To construct the
characteristic function let us redefine $\lambda \to (2 \pi)^{2-\psi}
\lambda /(|A|L^{\zeta-2} ) $, that corresponds to measuring $f_c$ in
units of $|A|L^{\zeta-2}/ (2 \pi)^{2-\psi}$. This is a non-universal
scale as the value of $\tilde{\Delta}^{*\prime}(0^+)$ is not universal
at the depinning transition.  However as we now show, once rescaled
the (shifted) distribution is universal.

The characteristic function can be written as
\begin{eqnarray}
\ln \hat{P}(\lambda)=-\frac12\sigma^2\lambda^2-
\sum\limits_{\mathbf{k} \in Z^d, \mathbf{k} \neq0}
\sum\limits_{n=3}^{\infty} \frac{1}{n}
\left(\frac{i\lambda}{|\mathbf{k}|^{2-\psi }} \right)^n, \label{eq66}
\end{eqnarray}
where
\begin{eqnarray}\label{sigma-r}
\sigma^2&=& \frac{(2\pi)^4 \tilde{\Delta}^*(0)}{ A^2 \tilde{I}_1
\varepsilon}  \nonumber \\
&=& \epsilon \tilde I_{1}\,  \frac{\tilde
\Delta^{*} (0)}{ [\tilde \Delta'^{*} (0)]^{2}}\, (2\pi)^{4-2\psi}=
\frac{6\pi^2}{\varepsilon}(1+\mathcal{O}(\varepsilon)).\qquad
\end{eqnarray}
To compute this universal ratio to lowest order in $\epsilon$ we have
used the 1-loop FRG fixed-point equation evaluated at $u=0$, i.e.\
$(\epsilon - 2 \zeta) \tilde{\Delta}^* (0) =
\tilde{\Delta}^{*\prime}(0^+)^2$ and used $\zeta=\epsilon/3 +
O(\epsilon^2)$, as well as $\epsilon \tilde I_1 = 1/(8 \pi^2) +
O(\epsilon)$.  Summing over $n$ we obtain
\begin{eqnarray}
 \ln \hat{P}(\lambda)&=&-\frac12\sigma^2\lambda^2 +
 \sum\limits_{\mathbf{k} \in Z^d, \mathbf{k} \neq0}
\left[ i\lambda \frac1{|\mathbf{k}|^{2-\psi}} \right. \nonumber \\
 &&- \left.\frac12 \lambda^2 \frac1{|\mathbf{k}|^{2(2-\psi)}}+ \ln
\left( 1- \frac{i\lambda}{|\mathbf{k}|^{2-\psi}} \right)
\right]. \qquad \label{eq67}
\end{eqnarray}

\subsection{Lowest order in $\epsilon=4-d$}
To obtain the distribution
within the $\epsilon$ expansion it is sufficient to set $\psi=0$ in
the formula above, and to compute all sums in $d=4$. Let us first give
the skewness and kurtosis to lowest order in the $\epsilon$
expansion. One uses \cite{ledoussal03}:
\begin{eqnarray}
&& \sum_{\mathbf{k} \in Z^d, \mathbf{k} \neq0} \frac1{|\mathbf{k}|^{2 p}} =
\frac{1}{(p-1)!} \nonumber \\
&& \ \ \ \ \ \ \ \ \ \ \ \ \ \ \ \ \ \
 \times \int_0^\infty dt\, t^{p-1} [\Theta(3,0,e^{-t})^d-1],  \\
&& \sum_{\mathbf{k} \in Z^4, \mathbf{k} \neq0} \frac1{|\mathbf{k}|^{6}} = 14.8298, \\
&& \sum_{\mathbf{k} \in Z^4, \mathbf{k} \neq0}
\frac1{|\mathbf{k}|^{8}} = 10.2454,
\end{eqnarray}
where $\Theta(3,0,e^{-t}) = \sum_{k \in Z} e^{- t k^2} $ is
the elliptic theta function. Hence:
\begin{eqnarray}
 \sigma_3 &=& \frac{\overline{(f-\bar{f})^3}}{\sigma^3}=\frac{2}{\sigma^3}
\sum\limits_{\mathbf{k} \in Z^d, \mathbf{k} \neq0} \frac1{|\mathbf{k}|^{6}} \\
& =&  0.0650861 \epsilon^{3/2},  \\
\sigma_4 &=& \frac{\overline{(f-\bar{f})^4}}{\sigma^4}-3 =-\frac{3!}{\sigma^4}
\sum\limits_{\mathbf{k} \in Z^4, \mathbf{k} \neq0} \frac1{|\mathbf{k}|^{8}} \\
&=& - 0.01753 \epsilon^{2}.
\end{eqnarray}
Next one can resum to obtain the characteristic function to lowest
order in $\epsilon$:
\begin{eqnarray}
 \ln \hat{P}(\lambda)&=&-\frac12\sigma^2\lambda^2 - F( - i \lambda),  \\
F( - i \lambda)    &=&
 \int_0^\infty \frac{dt}{t} (e^{i \lambda t}-1-i\lambda t+\frac{1}{2}
  \lambda^2 t^2)\nonumber \\
&& \qquad \qquad \times [\Theta(3,0,e^{-t})^4-1].
\end{eqnarray}
This result can be reexpressed as follows.
As $d \to 4^-$ the shifted dimensionless critical force $\tilde f = (f_c(L)-\overline{f_c(L)})/\sqrt{\overline{f_c(L)^2}^c}$ becomes
a univariate gaussian random variable. For small $\epsilon>0$, $\tilde f$
can be (formally) expressed as the sum of two independent random variables
$\tilde f = f_0 + \frac{\sqrt{\epsilon}}{\sqrt{6} \pi} f_1$ where $f_0$ is gaussian of variance
$1+O(\epsilon)$ and $f_1$ is a random variable of order unity with a non trivial
distribution, the logarithm of its characteristic function being given (up to a quadratic term)
by $F(- i \lambda)$.

We now analyze the shape of these distributions
in physical dimension.

\subsection{Fourier inversion}
In this section we compute the inverse
Fourier transform of Eq.~(\ref{eq67}) in physical dimensions, using
our improved scheme.

\begin{figure}[tbp]
\includegraphics[clip,width=2.8in]{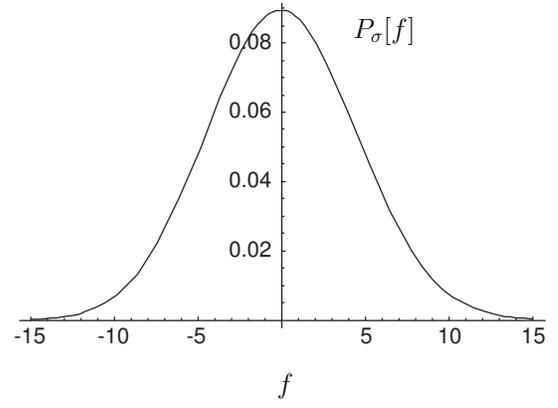}
\caption{The three indistinguishable curves for
shifted critical-force distribution ($d=1$): improved
perturbation theory ($\psi=0$), renormalized to one-loop theory
and approximate expression (\ref{eq70}). Here $\sigma=\pi\sqrt{2}$.
The difference between distributions is shown in Figure~\ref{fig3b}.
}
\label{fig3}
\end{figure}

\begin{figure}[tbp]
\includegraphics[clip,width=3.2in]{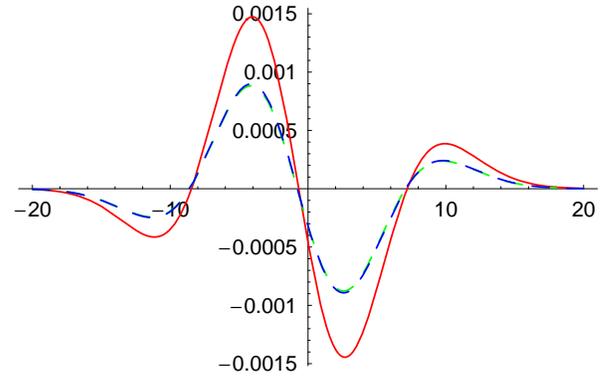}
\caption{(Color online) Shifted
critical-force distribution for $d=1$ with  Gaussian
subtracted. The dashed lines: approximate expression
(\ref{eq70}) and improved perturbation theory ($\psi=0$); solid line:
renormalized to one-loop theory. }
\label{fig3b}
\end{figure}

Let us start by discussing $d=1$. We use a natural extrapolation of
our above result,  setting $\epsilon=3$ in the above formulae (which
are exact to lowest order in $\epsilon$). From Eq.~(\ref{eq67}) we
obtain
\begin{eqnarray}
\hat{P}(\lambda) &=&
\exp\left[-\pi^2 \lambda^2 \right] \prod \limits_{k=1}^{\infty} \left\{ \left(
1-\frac{i\lambda}{k^{2/3}}\right)^2 \right.\nonumber \\
 &&\times \left. \exp\left[i\lambda \frac2{k^{2/3}} -  \lambda^2 \frac1{k^{4/3}} \right]
\right\}, \label{eq68}
\end{eqnarray}
where we have used $\psi=4/3$ and $\sigma=\pi\sqrt{2}$.  The inverse
Fourier transform computed numerically is shown in Figure~\ref{fig3}.
Eq.~(\ref{eq67}) with $\psi=0$ can formally be considered as the
result of improved perturbation theory in non-analytic disorder. The inverse Fourier
transform of the latter is also shown in Figure~\ref{fig3} and can not
be visually distinguished from the shifted renormalized
distribution. The renormalized distribution is more appealing since it
guarantees a non-negative defined critical force, which is not the
case for the bare one, especially in $d=1$ where the bare averaged
critical force (\ref{f-mean}) is finite. By contrast the averaged
renormalized critical force is controlled by the UV cutoff and is of
order $L_c^{-1/\nu}$ while the typical fluctuation is much smaller, of
order $L^{-1/\nu}$, in the limit of interest, here $L \gg L_c$.  It is
interesting that the renormalized distribution is  well
approximated by the Eq.~(\ref{eq68}) in which we keep only the first
factor with $k=1$:
\begin{eqnarray}
\hat{P}(\lambda)\approx
\exp\left[-\frac12\sigma^2\lambda^2 \right]  \left\{ \left(
1-{i\lambda}\right)^2
\exp\left[2i\lambda  -  \lambda^2  \right]
\right\}, \label{eq69}
\end{eqnarray}
The inverse Fourier transform of Eq.~(\ref{eq69}) reads
\begin{equation}
P[f]\approx  \frac {\left(
(\sigma^2+f+4)^2 - 2-\sigma^2
 \right)
}{\left( 2+{\sigma}^{2} \right) ^{5/2} \sqrt {2\pi }}
e^{-{\frac { \left( 2+f \right) ^{2}}{2(2+{\sigma}^{2}) }}}
\label{eq70}
\end{equation}
and is also shown in Figure~\ref{fig3}. The difference in
the critical force distributions obtained within
improved perturbation theory, renormalized to
one-loop theory and approximation (\ref{eq70}) is indicated
in Figure~\ref{fig3b}.

For a $d$-dimensional system the Fourier transform  of the critical force
distribution (\ref{eq67}) can be written as
\begin{eqnarray}
\hat{P}(\lambda)&=&\exp\left[-\frac12 \sigma^2 \lambda^2\right]
\prod\limits_{\mathbf{k} \in Z^d, \mathbf{k} \neq0}
\left\{
\left(1 - \frac{i\lambda}{|\mathbf{k}|^{2-\psi}}   \right)
\right. \nonumber \\
& \times & \left.\exp
\left[ i\lambda \frac1{|\mathbf{k}|^{2-\psi}}
-\frac12 \lambda^2
\frac1{|\mathbf{k}|^{2(2-\psi)}}
\right] \right\}.
\label{res4}
\end{eqnarray}
Analogously to the case $d=1$ the inverse Fourier transform
of Eq.~(\ref{res4}) can be well approximated by
\begin{eqnarray}
\hat{P}(\lambda)=
\exp\left[-\frac12\sigma^2\lambda^2 \right]  \left\{ \left(
1-{i\lambda}\right)^{2d}
\exp\left[2 d i\lambda  -  d \lambda^2  \right]
\right\}, \ \label{eq72}
\end{eqnarray}
which does not depend on $\psi$.


We now compute the standard deviation, and the kurtosis of the
above distributions.  The standard deviation reads
\begin{equation}\label{sd}
   \overline{(f-\bar{f})^2} \equiv  \sigma^2.
\end{equation}
The skewness is defined as
\begin{equation}\label{skewness}
  \sigma_3 = \frac{\overline{(f-\bar{f})^3}}{\sigma^3}=\frac{2}{\sigma^3}
\sum\limits_{\mathbf{k} \in Z^d, \mathbf{k} \neq0} \frac1{|\mathbf{k}|^{3(2-\psi)}}
\end{equation}
For $d=1$ we obtain $\sigma_3=\sqrt{2}/(6\pi)\simeq 0.075$ [$0.046$
for the distribution~(\ref{eq70})].  The positive value for the
skewness indicates that the right tail of the distribution is heavier
than the left tail.

The kurtosis for a Gaussian distribution
is three. For this reason, excess kurtosis is defined as
\begin{equation}\label{kurtosis}
  \sigma_4 = \frac{\overline{(f-\bar{f})^4}}{\sigma^4}-3 =-\frac{3!}{\sigma^3}
\sum\limits_{\mathbf{k} \in Z^d, \mathbf{k} \neq0} \frac1{|\mathbf{k}|^{4(2-\psi)}}
\end{equation}
For $d=1$ we obtain $\sigma_4=-3\zeta(8/3)/\pi^4 \simeq -0.04$
[$-0.031$ for the distribution~(\ref{eq70})]. Here $\zeta(x)$ is the
Riemann Zeta function. The small negative excess kurtosis indicates
that the distribution is slightly more flat than a Gaussian
distribution, while the deviation from the Gaussian distribution is
quite small.

\section{Periodic systems}
\label{sec3}

The renormalization of the critical force distribution for periodic
systems requires a separate consideration. Indeed as was shown in
Ref.~\cite{narayan-fisher93}, in the periodic case there is an
additional relevant operator which is the uniform part of
$\Delta(u)$ so that the RP fixed point is unstable. The flow equation
for this operator can be derived by integration of the RG equation
over one period \cite{ledoussal02}
\begin{equation}\label{extra}
  L\partial_L \int\limits_0^{1} \tilde{\Delta }(u) d u= \varepsilon
  \int\limits_0^{1} \tilde{\Delta }(u) d u + \mathcal{O}(\Delta^3),
\end{equation}
where we have explicitly used $\zeta=0$. Thus in the vicinity of the
RP FP, the flow of the dimensionless disorder is given by
\begin{equation}\label{extra2}
  \tilde{\Delta }(u)= \tilde{\Delta }^{*}(u) + c L^{\varepsilon}
\end{equation}
where the non-universal constant $c$ can be estimated as \cite{ledoussal02}
\begin{eqnarray}\label{constant-c}
  c &=& L_c^{-\varepsilon} \int\limits_0^{1} \left(\tilde{\Delta }^{\mathrm{(bare)}}(u) -
  \tilde{\Delta }^*(u)\right) d u \nonumber \\
  &=& - L_c^{-\varepsilon} \int\limits_0^{1}   \tilde{\Delta }^*(u) d u
  = L_c^{-\varepsilon} \left(\frac{\varepsilon^2}{108}+\mathcal{O}(\varepsilon)\right) >0.
  \nonumber
\end{eqnarray}
This runaway correction to the scaling behavior at the RP FP
contributes to all quantities which depend on $\Delta(0)$ but not to
those which depend on
$\Delta'(0)$. Therefore in the case of a periodic system the
renormalized second cumulant of the critical force reads
\begin{equation} \label{periodic}
\overline{\langle f_c(L)^2 \rangle}_c  = \frac1{\tilde{I}_1
 \varepsilon} \tilde{\Delta}^*(0)L^{-4} + \frac{c}{\tilde{I}_1\varepsilon}L^{-d}.
\end{equation}
Higher order cumulants are still given by Eq.~(\ref{f-n-ren}) with
$\zeta=0$ and therefore scale with $L$ as $\overline{\langle f_c(L)^n
\rangle}_c = L^{-2n}$.  We would like to emphasize that the correction
to scaling in Eq.~(\ref{periodic}) describes the sampel-to-sample
fluctuations and can not be seen in one sample because in each sample
there is only one pinned configuration.  As a result for $d<4$ only
the second cumulant of the {\em scaled} critical force, which scales
as $L^{-d}$, survives in the limit $L \to \infty $ resulting in
$\nu_{\mathrm{FS}}=2/d$ and a pure Gaussian distribution for the
scaled critical force.

\section{Discussion}
\label{sec4}

In the present paper we have computed the renormalized distribution
$P[f-\eta v]$ averaged over all pinned configurations in the limit $v
\to 0^{+}$, which we identify with the critical force distribution
$P_L(f_c)$. The average critical force is a non-universal quantity
which depends on microscopic details of the interactions like the UV
cutoff as well as on details of the disorder distribution. After
subtraction of the average value, the shifted distribution of the
critical force contains only one non-universal scale which can be
fixed, e.g.~by fixing the second cumulant. The resulting dimensionless
distribution is then fully universal, i.e.\ it does not depend on
properties at small scales. We have computed it taking into account
only the second cumulant of the bare disorder distribution, since it
is the only cumulant relevant in the RG sense. Higher cumulants, which
are generated by coarse graining are irrelevant operators and their
contribution to the cumulants of the critical force must carry
additional dependence on the cutoff. Hence we expect that they result
only in a shift of the non-universal expectation values.

Let us now discuss the role of the transverse sample size $M$ (size of
the box). In numerical studies of depinning of elastic interfaces,
either via an exact determination of the critical state, or via
Langevin dynamics \cite{rosso03,bolech04,Olaf05,rosso2003,rosso2001} a
cylindrical system which is periodic in both directions was studied:
longitudinal with period $L$ and transverse with period $M$. Hence
this is equivalent to a periodic disorder with period $M$. It is known
that a periodic system has a unique pinned configuration for any
period $M$ \cite{middleton92}. If $M\ll L$, this configuration spreads
out through the whole box $L^d \times M$ and there is only one
independent pinned configurations. As we have shown for the RP class
the distribution of critical forces is Gaussian, and thus, for elastic
interfaces in the limit $L \to \infty $, with $M$ fixed, the
distribution of the critical force also becomes Gaussian.

The case where the period $M$ is taken to infinity at the same time as
$L$ is relevant for elastic interfaces and quite different. The pinned
interface has a r.m.s. width $w=k_w L^{\zeta}$ so that in a sample
of transverse size $w$ it also has one unique statistically independent
pinned configuration. One may then argue that its critical force
distribution is $P_L(f_c)$ whose characteristic function is given by
Eq.~(\ref{eq67}). In the numerical studies one should thus be careful
in choosing the size of the periodic box $M$.  If $M$ scales like
$L^{\zeta'}$ with $\zeta'<\zeta$ the system will crossover from the
random manifold to the RP FP and the finite-size scaling analysis will
results in some mixture of interface and periodic system properties,
while the critical force distribution will tend to a Gaussian one.  On
the over hand if $M \gg w $, the sample can be divided into about
$M/w$ subsamples, which can be argued to be (almost) statistically
independent, with independent pinned configurations.  Each
configuration has a slightly different critical force which is
distributed according to our FRG result. If one defines the total
critical force as a maximum of all the critical forces of these
subsamples, it becomes $M$-dependent and its shifted distribution
tends to the distribution of the extreme value statistics
\cite{bolech04}.  Following Ref.~\cite{bolech04} let us introduce for
every configuration $\alpha$ of the interface in the sample of width
$M=k L^{\zeta}$ the depinning force $f_d(\alpha)$ and then associate
the threshold force of the whole sample with the following maximal
value $ f^{r}_c=\max_{\alpha} \left\{f_d(\alpha)\right\} $.  In each
sample there are only $\approx M/w$ independent pinned configuration,
so that the distribution of the maximum of the corresponding critical
forces can be written as
\begin{equation}\label{Pm}
  P_M(f^{r}_c, M/w)=\frac{d}{d f^r_c} \left[\int\limits_{-\infty}^{f^{r}_c} df' P_L[f'],
  \right]^{M/w}\ .
\end{equation}
According to the general theorem of extreme value statistics
\cite{galambos87}, for large samples, i.e.\ , in the limit $M/w \to
\infty$ this distribution approaches the Gumbel distribution. The
latter is provided by the {\em tails} of the distribution of the critical
force for each independent pinning configuration, as given in
Eq.~(\ref{eq67}).  According to this distribution the average maximal
threshold force of samples of size $M$ behaves as $\ln(k/k_w)$. For
large samples with $M \gg w$ it can be extremely large.  The above
procedure completely washes out all details of the underlying
distribution $P_L(f)$ computed here, except for its width, and
replaces it by the model-independent function obtained from
extreme value statistics. As an illustration, we have plotted the
force distribution obtained using (\ref{Pm}) for $M/w=10$ on figure
\ref{}.

\begin{figure}[tbp]
\includegraphics[clip,width=3.2in]{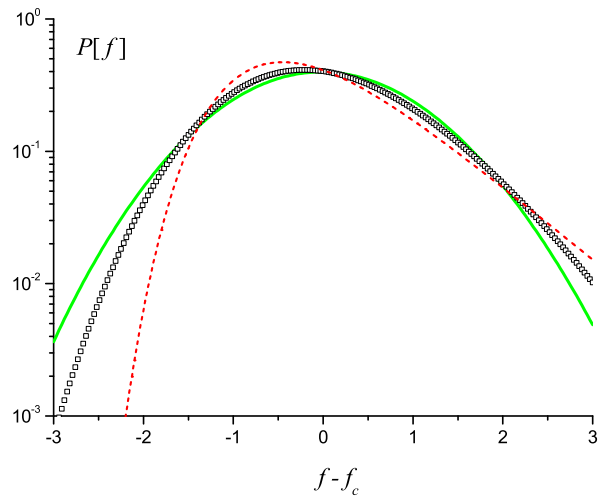}
\caption{ (Color online)
The normalized critical-force distribution for $d=1$.  The solid line is
the distribution for the interface in the box of size $w$ given by
Eq.~(\ref{eq70}). The dashed line is the Gumbel distribution, i.e.\
the distribution of the maximal threshold force in the limit of an
infinite box.  The points are computed using Eq.~(\ref{Pm}) for the
interface in the finite box of size $M=10w$.}
\label{fig4}
\end{figure}

The above arguments suggest that the critical force distribution
computed here via the FRG should be compared with the numerics on a
cylinder of aspect ratio parameter $k \approx k_w$ defined above from
the width. We now propose a more precise statement to identify the
critical force computed in this paper. We note that in the
calculations performed here within the FRG the position of the center
of mass was held fixed (since all momentum integrations excluded the
uniform mode $q=0$). Hence we are working in the {\it fixed center of
mass ensemble}. This suggests the definition:
\begin{equation}
f_c(u_0,L) = \max_{\alpha(u_0,L)} \left\{f_d(\alpha(u_0,L))\right\}
\label{deffc}
\end{equation}
where the maximum is over all configurations $\alpha(u_0,L)$ with
center of mass $u_0$ and length $L$ (and periodic boundary conditions
along the interface). It can in principle be evaluated numerically by
direct enumeration for a discrete interface model. One can then check
that it has a well-defined $L \to \infty$ limit with no need for a
transverse box, and one can then numerically compute the finite-size
distribution for the ensemble of $\alpha(u_0,L)$. This distribution
should identify with the one computed here within the FRG (in the
massless scheme). It is a more fundamental object than the critical
force defined on a cylinder. The latter can then be retrieved in
principle as
\begin{equation}
f^{r}_c = \max_{u_0} f_c(u_0,L)
\end{equation}
on the same cylinder, leading to extremal statistics, as discussed
above.

The above considerations illustrate that the statistics of the
depinning threshold force at finite-size is a rather subtle
question. Many questions remain open. It would be interesting to find
the proper steady state corresponding to the above definition
(\ref{deffc}). Also a systematic study of memory effects in the
threshold force would be of high interest especially regarding
experiments. Indeed, these memory effects may be of importance for
aging \cite{vincent-00} and hysteresis phenomena \cite{nattermann-01}
controlled in some regimes by the slow dynamics of domain walls. There
the observed threshold force may not be the largest one but a
threshold force which characterizes a piece of the sample in which the
interface got trapped. Hence the data should be interpreted with care
to disentangle history effects from finite size effects. It would be
very interesting to develop numerical schemes to investigate these
questions in particular an efficient algorithm to compute
(\ref{deffc}).

 \begin{acknowledgments}
We thank O. Duemmer, W. Krauth, A. Rosso for useful discussions.
AAF acknowledges the support from the European Commission through
Marie Curie Postdoctoral Fellowship under contract number MIF1-CT-2005-021897.
PLD and KJW acknowledge support from ANR program 05-BLAN-0099-01.
\end{acknowledgments}

\appendix

\section{Calculation of the asymptotics of $\Gamma^{(+)}_{\hat{u}\hat{u}u}$}
\label{appendix}

To renormalize the $n$-th cumulant of the critical force we need the
leading asymptotics of the vertex functions
$\Gamma^{(+)}_{\hat{u}\hat{u}u}$ and
$\Gamma^{(-)}_{\hat{u}\hat{u}u}$. The latter is insensitive to the
IR cutoff scheme up to corrections to the amplitude. In order to
extract this asymptotics, we adopt a massive scheme for the IR
regularization because it leads to the simplest calculations. If one
then needs the corresponding amplitude in a different scheme, one
can relate it to the amplitude in the massive scheme following the
methods developed in Ref.~\cite{ledoussal04}.

Straightforward perturbation theory gives to first order in the bare
disorder denoted here $\Delta_0$ for the ``open'' vertex
$\Gamma_{\hat{u}\hat{u}u}^{(+)}$
\begin{eqnarray}
&&\int_{t_2}\Gamma_{\hat{u}\hat{u}u}^{(+)}(t,t_1,t_2 ;q,-q)=
\Delta'_0(0^+)\left\{1-\Delta''_0(0) \right. \nonumber \\
&& \ \ \ \ \ \ \ \ \ \  \times  \int_p \left[ \frac{2}{(p^2+m^2)[(p+q)^2+m^2]}
                       \nonumber \right. \\
&& \ \ \ \ \ \ \ \ \ \ +\left.\left.\frac{1}{(p^2+m^2)^2} +
     O \left( e^{-p^2(t-t_1)} \right ) \right] \right\}. \label{delta0}\\
&&\int_{t_2}\Gamma_{\hat{u}\hat{u}u}^{(-)}(t,t_1,t_2 ;q,-q)=-
\int_{t_2}\Gamma_{\hat{u}\hat{u}u}^{(+)}(t,t_1,t_2 ;q,-q)\qquad~. \label{delta0-2}
\end{eqnarray}
Identity (\ref{delta0-2}) holds to all orders by definition. The
last term in Eq.~(\ref{delta0}) reflects the dynamic nature of the
vertex $\Gamma_{\hat{u}\hat{u}u}^{(+)}(t,t_1,t_2 ;q,-q)$.  However,
if we integrated this term also over $t_1$ the result will not
depend on the observation time $t$. As a consequence, the n-th
cumulant of the critical force  is determined by the quasi static
integrals (to any order in disorder).

\begin{figure}[tbp]
\centerline{\includegraphics[width=4cm]{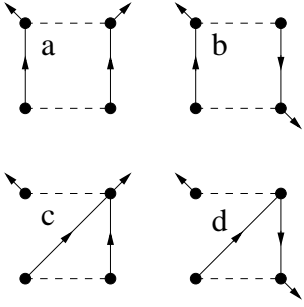}}
\caption{1-loop dynamical diagrams correcting $\Delta$.}
\label{1loopDelta}
\end{figure}

The easiest way to obtain (\ref{delta0}) is as follows: One starts
from the 1-loop diagrams given in Figure \ref{1loopDelta}, expands
to first order in the fields $u$, and then replaces $\Delta_0^{(n)}
(u_{t}-u_{t'})$ by $\Delta_0^{(n)} (0^{+})\times [ \mbox{sign}
(t-t')]^{n}$. The diagrams are then split into the two classes
$\Gamma^{(-)}_{\hat{u}\hat{u}u}$ and
$\Gamma^{(+)}_{\hat{u}\hat{u}u}$, depending on whether the single
field $u$ is connected to $\hat{u} (t_{1})$ or $\hat{u} (t_{2})$. In
order to extract the time-independent terms, one choses $t-t_{1}\to
\infty$. This prescriptions allows for an easy integration of all 16
diagrams. Denoting
\begin{equation}
I_1(q)=\int_p
\frac{1}{(p^2+m^2)[(p+q)^2+m^2]}=m^{-\varepsilon}\tilde{I}_1(q/m)\ ,
\end{equation}
and $\tilde I_1=\tilde I_1(0)$, the contributions to (\ref{delta0})
are as follows
\begin{eqnarray}\label{lf1}
&&- \Big\{[I_{1}(q)] + [ I_{1}(0)+ I_{1}(q)] +[0] + [ I_1(0)-I_{1} (0)]
\Big\}\nonumber \\
&&\qquad \times  \Delta_0 '(0) \Delta_0 ''(0)\nonumber \\
&& +\Big\{ [- {I_1}(0)] + [0]+[ {I_1}(0)]+[0]\Big\} \Delta_0 (0),
\Delta_0'''(0)\qquad
\end{eqnarray}
where terms in rectangular brackets are in the order of their
appearance from diagrams a, b, c and d of figure \ref{1loopDelta}.
We remark that contributions proportional to $\Delta_0 (0)
\Delta_0''' (0)$ exactly cancel, and we obtain (\ref{delta0}).

To renormalize the vertex function (\ref{delta0}) we have to reexpress
the bare disorder correlator by the renormalized dimensionless one:
\begin{eqnarray}\label{}
&&\!\!\!\Delta_0(u)=\nonumber \\
&&m^{\varepsilon} \left \{ \Delta(u) + \left[ \Delta'(u)^2 +
(\Delta(u)-\Delta(0)) \Delta''(u) \right] \tilde I_1 \right
\}.\nonumber
\end{eqnarray}
Differentiating the latter expression w.r.t. $u$ we get after
rescaling $\Delta(u)= \frac{1}{\varepsilon\tilde{I}_1}
m^{-2\zeta}\tilde{\Delta}(um^{\zeta})$
\begin{eqnarray}
\Delta'_0(0) &= & \frac{m^{\varepsilon-\zeta}
}{\varepsilon\tilde{I}_1} \tilde{\Delta}'(0^+)\left[1+
\frac{3}{\varepsilon\tilde{I}_1} \tilde{\Delta}''(0)m^{\varepsilon}
\int_p  \frac{1}{(p^2+m^2)^2} \right], \nonumber \\
\Delta''_0(0) &= & \frac{1}{\varepsilon\tilde{I}_1} m^{\varepsilon}
\tilde{\Delta}''(0^+) + \mathcal{O}\left(\tilde{\Delta}^{\prime\prime}(0)^2,
 \tilde{\Delta}^{\prime}(0^+)\tilde{\Delta}^{\prime\prime\prime}(0^+) \right ). \nonumber
\end{eqnarray}
Omitting the last term in Eq.~(\ref{delta0}) we obtain the following
expression for the renormalized vertex function
\begin{eqnarray}
&& \mbox{} \hspace{-15mm} \int_{t_2}
\Gamma_{\hat{u}\hat{u}u}^{(+)}(t,t_1,t_2 ;q,-q)= m^{\varepsilon-\zeta}
\frac{1}{\varepsilon\tilde{I}_1}\tilde{\Delta}'(0^+) \nonumber \\
&& \times \left[1- 2\tilde{\Delta}''(0)\frac{1}{\varepsilon\tilde{I}_1} m^{\varepsilon}
 [I_1(q)-I_1(0)] \right], \nonumber
\end{eqnarray}
Note that depinning of the non-periodic systems is described by the
fixed point with
$\tilde{\Delta}''(0)^*=\frac{2}{9}\varepsilon+O(\varepsilon^2)$. The
one-loop integral $\tilde{I}_1(y)$ reads \cite{ledoussal02}
\begin{equation}
\tilde{I}_1(y)= \frac12
K_d\Gamma\left(\frac{d}2\right)\Gamma\left(\frac{\varepsilon}2\right)
\int\limits_0^{1} \frac{d
\alpha}{[1+\alpha(1-\alpha)y^2]^{\varepsilon/2}},
\end{equation}
where $K_d=2\pi^{d/2}/((2\pi)^d\Gamma(d/2))$ is area of a $d$
dimensional sphere divided by $(2\pi)^d$.
 Taking into account that
$\tilde{I}_1\equiv\tilde{I}_1(0) =\int_q (q^2+1)^{-2}=\frac12%
K_d\Gamma\left(\frac{d}2\right)\Gamma\left(\frac{\varepsilon}2\right)$
we obtain
\begin{eqnarray}
&&\int_{t_2}\Gamma_{\hat{u}\hat{u}u}^{(+)}(t,t_1,t_2 ;q,-q) =
m^{\varepsilon-\zeta}
\frac{1}{\varepsilon\tilde{I}_1}\tilde{\Delta}'^*(0^+)
\nonumber \\
&& \qquad \times
\left[1+ \frac29 \varepsilon \int\limits_0^{1}
d \alpha  \ln \left[1+\alpha(1-\alpha)y^2\right]+ O(\varepsilon^2) \right].
\nonumber
\end{eqnarray}
We are interested in the asymptotic behavior for $z \to \infty$. In
this limit we have
\begin{eqnarray}
&&\int\limits_0^{1}
d \alpha  \ln \left[1+\alpha(1-\alpha)y^2\right]=-2+\frac{\sqrt{4+y^2}}{y}\Big[
\ln 2  \nonumber \\
&& -  \ln(2+y^2-y\sqrt{4+y^2})
\Big] = -2+2\ln y + O\left(\frac{\ln y}{y^2}\right). \nonumber
\end{eqnarray}
Matching to a power-law asymptotic behavior we find
\begin{equation}\label{}
1+ \frac29 \varepsilon (-2+ 2\ln y)  + \mathcal{O}(\varepsilon^2)  \to
y^{4\varepsilon/9}\left(1-\frac49 \varepsilon\right).
\end{equation}
As a result we obtain for $q/m \gg 1$:
\begin{eqnarray}
&&\mbox{} \hspace{-15mm} \int_{t_2}\Gamma_{\hat{u}\hat{u}u}^{(+)}(t,t_1,t_2 ;q,-q)
\nonumber \\
&&= m^{\varepsilon-\zeta}
\frac{1}{\varepsilon\tilde{I}_1}\tilde{\Delta}'^*(0^+)
\left(\frac{q}{m} \right)^{4\varepsilon/9} \left(1-\frac49 \varepsilon\right).
\end{eqnarray}
Replacing $m$ by $1/L$ we obtain Eq.~(\ref{Gamma3}).


\end{document}